# Improvement of Morphology and Electrical Properties of Boron-doped Diamond Films via Seeding with HPHT Nanodiamonds Synthesized from 9-Borabicyclononane

Stepan Stehlik[1,2]*, Stepan Potocky[1,3], Katerina Aubrechtova Dragounova[1], Petr Belsky[2], Rostislav Medlin[2], Andrej Vincze[4], Evgeny A. Ekimov[5], Alexander Kromka[1]

[1] Institute of Physics, Czech Academy of Sciences, Prague, Czech Republic

[2] New Technologies Research Centre, University of West Bohemia, Pilsen, Czech Republic

[3] Faculty of Electrical Engineering, Czech Technical University in Prague, Prague, Czech Republic

[4] International Laser Centre, Slovak Centre of Scientific and Technical Information, Bratislava, Slovak Republic

[5] Vereshchagin Institute for High Pressure Physics, Russian Academy of Sciences, Troitsk, Russia

*Corresponding author: stehlik@fzu.cz

**Abstract**

Boron-doped diamond (BDD) films are becoming increasingly popular as electrode materials due to their broad potential window and stability in harsh conditions and environments. Therefore, optimizing the crystal quality and minimizing defect density to maximize electronic properties (e.g. conductivity) of BDD is of great importance. This study investigates the influence of different hydrogenated nanodiamond (H-ND) seeding layers on the growth and properties of BDD films. Three types of seeding H-NDs were examined: detonation (H-DND) and top-down high-pressure high-temperature NDs (TD_HPHT H-ND), and boron-doped NDs (H-BND) newly synthesized at high-pressure high-temperature from an organic precursor. Purified and oxidized BND (O-BND) samples yielded clear, blue, and stable colloidal dispersions. Subsequent thermal hydrogenation reversed their zeta potential from – 32 mV to + 44 mV and promoted the seeding of negatively charged surfaces. All three H-ND types formed dense seeding layers on $SiO_2$ and $Si/SiO_x$ substrates, which enabled the growth of BDD films by chemical vapor deposition (CVD). Despite variations in initial surface coverage among the seeding layers (13-25%), all NDs facilitated the growth of fully closed BDD films approximately 1 μm thick. Significant differences in film morphology and electrical properties were observed. H-BND nucleation yielded the BDD films with the largest crystals (up to 1 000 nm) and lowest sheet resistance (400 ohm/sq). This superior performance is attributed to the uniform particle shape and monocrystalline character of H-BND, as corroborated by FTIR, TEM, and SAXS measurements. These findings highlight the critical role of seeding layer properties in determining consequent diamond film evolution and establish H-BNDs as promising seeding material for the growth of high-quality BDD films suitable for electronic and electrochemical applications.

**Introduction**

Recent advances in chemical vapor deposition (CVD) bring the industrial use of synthetic diamond layers closer. Here, boron-doped polycrystalline diamond films (BDD) are gaining attention due to their enhanced electrical conductivity and potential applications in electronics, electrochemistry and sensors[1–4]. The electrical conductivity of the CVD films is primarily determined by the boron doping level via controlling the B/C ratio (ppm) in the gas mixture[5]. However, not only the gas compositions, but also the nucleation method applied to non-diamond substrates significantly impacts the diamond crystalline characteristics, including grain size, defect density, preferential orientation, and boron incorporation into the diamond bulk[6,7]. These factors ultimately influence the electronic properties of diamond thin films.

Among the diverse nucleation techniques available, seeding with diamond nanoparticles stands out as the most widely adopted method[8–10]. This approach, in its simplest form, approximates homoepitaxial growth on diamond seeds resulting in the development of continuous film over non-diamond substrates. The seeding process plays a pivotal role in determining the subsequent growth dynamics and ultimate characteristics of the diamond film, underscoring its significance in the overall CVD process[11].

Epitaxial growth involves the deposition of a crystalline material on the well-defined surface of a crystalline substrate, with the overlayer having the same crystalline orientation as the substrate. The crystallographic quality and cleanness of the substrate are important parameters that have a great impact on the crystalline quality of the grown material[12,13]. Microwave plasma-enhanced chemical vapor deposition (MPCVD) growth of diamond is typically performed on a nanodiamond-seeded substrate where the nanodiamond (NDs) seeds act as nuclei, and the diamond phase grows directly on them in a homoepitaxial regime. The seeding problematics has recently been reviewed[14].

Detonation nanodiamond (DND) is commonly used for substrate seeding among various types of nanodiamonds[15]. DNDs are relatively cheap, readily commercially available, and offer a wide range of surface modification options[16]. This is of great importance because surface modifications of DNDs allow to control magnitude and polarity of the zeta potential, thus achieving the desired electrostatic attraction between DNDs and different substrates such as Si[17], AlN[18,19], BN[20]. Substrate modifications[21] including the effect of hydrocarbon contamination have also been studied[22]. Many substrates, such as quartz glass or Si, have a negative surface zeta potential, and therefore hydrogenated DNDs with a positive zeta potential are often used to create dense and homogeneous seeding layers[17,23]. DND´s size reduction and ionic strength control can maximize the seeding density allowing the growth of ultra-thin (< 10 nm) MPCVD diamond films[9,11,24]. However, DNDs are highly defective internally polycrystalline diamond nanoparticles, that often lack clear facets[25]. Moreover, the radical nature of the thermal hydrogenation treatment of NDs connected with the desorption of the original surface functional groups[26,27] leads to inevitable surface reorganization and to overall increase of the surface disorder[28]. Thus, hydrogenated DND (H-DND) cannot be considered as a substrate with a high degree of crystal perfection and surface cleanness.

Another ND type that could be used as seeds is commercially produced by milling diamond microcrystals synthesized by the high-pressure high-temperature (HPHT) process. These top-down HPHT NDs (TD_HPHT NDs) are monocrystalline and provide much better crystal quality, chemical, and phase purity than DNDs

[29]. On the other hand, TD_HPHT NDs have rather broad size distribution as well as highly irregular shapes, resembling shards of glass[30]. Their single crystalline nature and generally larger size than DNDs make them less reactive to hydrogen[31,32]. Only recently the successful hydrogenation of TD_HPHT NDs and the preparation of stable colloids with a high enough positive zeta potential has been demonstrated[33–36]. This progress opens ways for using hydrogenated TD_HPHT NDs (TD_HPHT H-NDs) for seeding of common negatively charged substrates such as Si or silica. Indeed, a recent study[37] demonstrated that the crystalline quality of silica/diamond core-shells grown by MPCVD depends significantly on the nature of diamond seeds that initiate the growth. The use of TD_HPHT H-NDs instead of H-DND resulted in highly crystalline diamond coatings with a columnar growth at low methane concentration (0.4 vol %). Yet, similarly to DND, the thermal hydrogenation of TD_HPHT NDs leads to surface rearrangement, namely an increase of the non-diamond $sp^2$ carbon as well as the appearance of trans-polyacetylene chains[28,33] which compromises the surface cleanliness.

Obviously, the ideal nanodiamond seed would be a monocrystalline diamond nanoparticle with a regular shape and a clean, hydrogenated surface free of $sp^2$ C. Such NDs can indeed be produced by bottom-up HPHT synthesis from molecular precursors such as halogenated adamantanes[38–41]. The great advantage of the BU_HPHT NDs is that their surface hydrogenation is formed directly during the synthesis and no subsequent heat treatment is needed. As synthesized, BU_HPHT NDs have a fully hydrogenated surface with negligible $sp^2$-C content [42]. Unfortunately, the dispersion of BU_HPHT NDs in water and the formation of stable colloids that would be suitable for seeding is problematic and requires further effort[36]. Nevertheless, the bottom-up HPHT synthesis is quite a versatile technique and allows not only an unprecedented control over ND size[38] but also doping with Si and Ge for the formation of silicon-vacancy and germanium-vacancy color center[43,44], or with B for electrical conductivity[45] or optical absorption[46]. Interestingly, the few nm boron-doped ND (BND) synthesized from another organic molecular precursor, 9-borabicyclo[3,3,1] nonane dimer (9BBN), $C_{16}H_{30}B_2$[45] have been shown to be easily dispersible in water after an oxidative treatment[46]. BNDs synthesized in this way outperform CVD-grown BNDs[47], especially in terms of size and size distribution and yield available per synthetic run. This makes the BNDs highly attractive for seeding purposes owing to their few-nm size, monocrystalline character, and distinct faceting[48].

In this study, we first investigate the effect of thermal hydrogenation on the structural and colloidal properties of BNDs and their use as seeds along with the referential H-DND and TD_HPHT H-NDs. Finally, we evaluate the effect of the seed properties on the morphology, structure, and sheet resistance ($R_s$) of BDD films grown by MPCVD.

**Materials and Methods**

*BND synthesis and hydrogenation*

The BND samples were synthesized by the HPHT method from the 9BBN precursor (9-borabicyclo[3.3.1]nonane, $C_{16}H_{30}B_2$) at a pressure of 8.5-9 GPa and temperature of 1250 °C[45,48]. The dwelling time at constant pressure–temperature (P–T) parameters was 120 s. The synthesized dark powder was cleaned by boiling in a 3:1 mixture of sulfuric and nitric acids for 6 h until no bubbling was observed, yielding oxidized BND (O-BND). The O-BNDs were then washed by a multi-step centrifugation and replacing the acidic supernatant repeatedly by DI water at 20 000 g for 60 min (Sigma 3-30KS centrifuge) until the O-BNDs were no longer sedimented. The pH value of such obtained O-BND stable suspension was 4-5. The suspension was then lyophilized, and a fine powder was obtained. The hydrogenation of O-BND powder was performed in a tube furnace at an atmospheric hydrogen pressure of 700°C for 3 h providing the hydrogenated BND (H-BND). The 29 wt.% weight loss after hydrogenation indicated sub-5 nm size of H-BNDs, since an identical treatment of an oxidized DND (≈ 5 nm) resulted in only 11-12 wt.% loss.

*Preparation of seeding solutions and seeding layers*

H-BND powder (16 mg) was dispersed in 8 ml DI water (2 mg/ml) by sonication using a Hielscher UP200S ultrasonic processor for one hour at 120 W, 0.5 s on/off working cycle. DND solution was prepared by diluting the commercial NanoAmando single-digit DND dispersion (25 mg/ml, Nanocarbon Research Institute, Japan) from the original concentration to 1 mg/ml. The top-down HPHT H-ND (TD_HPHT H-ND) seeding solution was prepared using the commercially available monocrystalline HPHT NDs obtained by milling (MSY 0-0.03, Pureon). These were first purified from a residual $sp^2$-C by air annealing at 450°C for 5 hours and then hydrogenated at 800°C for 6 hours in a tube furnace at an atmospheric pressure of the hydrogen gas[28,33,36]. TD_HPHT H-ND powder (40 mg) was dispersed in 20 ml DI water (2 mg/ml) by sonication using the same equipment and conditions as described above. The obtained TD_HPHT H-ND dispersion was shortly centrifuged to remove the poorly dispersed particles at 14000 g for 10 minutes. The supernatant was carefully extracted by a micropipette. The final concentration was about 1.5 mg/ml.

The seeding layers were deposited on Si(100 orientation, dimensions: 5 x 10 x 0.53 mm) and fused $SiO_2$ substrates (5 x 10 x 1.4 mm). Before the deposition, the substrates were stepwise cleaned in acetone, isopropyl alcohol and deionized water in an ultrasonic bath for 10 minutes in each liquid and dried by nitrogen flow. Substrates were immersed in 1 ml Eppendorf tubes with the seeding solutions and sonicated for 10 minutes in a sonication bath (37 kHz). Subsequently, the substrates were shortly rinsed in DI water, and gently dried with nitrogen. Finally, the seeding layer was removed locally by gently wiping the substrate to enable direct measurement of the thickness of the BDD films by atomic force microscopy (AFM)[49].

*MPCVD growth and surface treatment of BDD films*

The CVD deposition of BDD films was performed in a microwave plasma-enhanced multimode clamshell cavity reactor system (SEKI SDS6K, Cornes Technologies, Ltd.). The constant process parameters were 5 % $CH_4$ of the $H_2/CH_4$/TMB gas mixture (TMB: trimethylboron), B/C ratio 5000 ppm, a total gas flow rate of 250 sccm, a chamber pressure of 8 kPa (60 Torr), substrate temperature ≈500 °C, and the CVD growth time 1 h. BDD growth on both substrates (Si and $SiO_2$) was carried out in one deposition run. To ensure hydrogen termination, the last step of the deposition was hydrogen plasma treatment for 10 minutes in the $H_2$ atmosphere. The other process parameters were kept the same. For the surface oxidation of BDD layers, an ICP radio frequency (RF, 13.56 MHz) plasma system with the following parameters was used: RF power 100 W, pressure 60 Pa, $O_2$ flow rate 25 sccm, and process time 2 min.

*Characterization methods*

Fourier transform infrared spectroscopy (FTIR) of O-BNDs and H-BNDs was measured by Thermo Nicolet iS50 spectrometer equipped with the KBr beam splitter and $N_2$-cooled MCT detector. 50 µL of the BND colloidal solutions was dropcasted on Au-coated Si substrates and shortly heated at 100 °C for 60 s to evaporate the liquid. The measurement chamber was continuously purged with dried air, ensuring very low water vapor pressure. All spectra were measured by a specular-apertured grazing angle reflectance (SAGA) method with an 80° incident light. Each spectrum represents an average of 128 scans per spectrum collected in the range of 4000–1000 cm$^{-1}$ at 4 cm$^{-1}$ resolution.

Raman spectra were acquired by a Renishaw InVia Reflex confocal system (Renishaw, UK) equipped with a 442 nm excitation laser (100 mW Dual Wavelength He–Cd laser, model IK5651R-G, Kimmon Koha), holographic grating 2400 lines/mm, and air-cooled CCD camera. For the measurements we used a Leica objective with 100× objective magnification (NA = 0.9). Before the measurements, the Raman spectrometer was calibrated on the crystalline diamond peak position (1332 cm$^{-1}$). The BND samples were deposited from the ND colloidal solutions on Si substrates by dropcasting and drying. Each sample was measured at 3 randomly chosen positions, where each of them was exposed by a continuous wave laser with a power of 0.4 mW focused to a spot of 1 μm for 100 s, and spectra were accumulated twice. Also, BDD film samples were measured at 3 different positions, using the same setup, only the laser was operated at a power of 40 mW, each spectrum was measured for 30 s, and spectra were accumulated three times. We first verified that such a combination of parameters did not damage the sample. After that, average spectra were calculated, and the baseline was subtracted from collected spectra (asymmetric least squares method: Asymmetric factor 10$^{-4}$, Threshold 0.003, Smoothing factor 5, 100 iterations) and Gaussian functions were used for peak analysis.

Dynamic light scattering (DLS) was measured using a Zetasizer Nano ZS (Malvern Panalytical) equipped with a helium-neon laser (633 nm). A scattering angle of 173° was used for all the measurements. The refractive index of bulk diamond (2.4) and the viscosity of pure water (0.89004 mPa.s at 25°C) were used. The size distribution measurements were done in polystyrene disposable cuvettes. Every sample was measured three times, and each of the three DLS size measurements consisted of 10 runs lasting 10 s. Zeta potential (ζ) measurements were performed in a disposable zeta potential measurement cell. A mean zeta potential value was calculated automatically in the software from the Henry equation after the estimation of the electrophoretic mobilities of the particles under the applied voltage in the zeta potential measurement cell. The zeta potential value was calculated as the average from 3 zeta potential values. All the samples were measured within one day after the preparation, the concentration of all ND dispersions was kept at 0.5 - 1 mg/ml for both size distribution and zeta potential analyses.

Small angle X-ray scattering (SAXS) measurements were conducted on a SAXSess mc$^2$ instrument (a Kratky-type instrument by Anton Paar) equipped with a microfocus X-ray source and single-reflection X-ray optics by Xenocs. The BND powder samples were glued between two pieces of scotch tape and SAXS patterns were recorded in the transmission mode with an exposure time of 10 min using a rectangular imaging plate (storage phosphor screen) as detector. The read-out of the imaging plate was carried out on a Cyclone® Plus Storage Phosphor System (PerkinElmer). The obtained 2D scattering patterns were first azimuthally averaged using the software supplied with the instrument, resulting in 1D scattering profiles. The 1D profiles were further processed with *Irena* software package [50], including subtraction of background scattering of the scotch tape and the incoherent scattering background and logarithmic data reduction. Finally, the Guinier analysis in the *Simple Fits and Analysis* tool and the *Unified Fit* tool in *Irena* [51] were applied to evaluate the morphological parameters of ND particles/clusters/aggregates.

High-resolution transmission electron microscopy (HRTEM) was carried out on a transmission electron microscope JEOL JEM 2200FS operated at 200 kV (Schottky field-emission gun, point resolution 0.19 nm) with an in-column energy Ω-filter for EELS/EFTEM, a STEM unit, and an Energy Dispersive X-ray (EDX) SDD detector Oxford Instruments X-Max attached. Images were recorded on a Gatan CCD camera with a resolution of 2048×2048 pixels using the Digital Micrograph software package. For HRTEM analysis, the TEM grids were briefly immersed in the respective BND colloid (oxidized and hydrogenated) and let dry prior to introduction into the TEM chamber.

The seeding layers and the surface morphology of BDD films were inspected by scanning electron microscopy (SEM, e_LiNE writer, Raith GmbH). The SEM images were further processed to obtain a total boundary length as an expression of the defectiveness of the BDD films grown on the H-DND, TD_HPHT H-ND and H-BND seeding layers. First, grains and grain boundaries were highlighted using ImageJ software (function: Find edges), and the processed images saved were further analyzed in Gwyddion software (function: Mark by Edge Detection). A threshold of 30% was used for all images and then the total boundary length values were obtained.

Secondary ion mass spectrometry (SIMS) depth profiles were acquired by using a Time-of-Flight SIMS IV (Ion-TOF GmbH) equipped with a $Bi^+$ primary ion gun working at 25 keV in dual beam mode. The sputtering was realized with $Cs^+$ ion guns at 2 keV using an active electron flooding to prevent the surface charging to increase the yield of negatively charged secondary ions. While the first beam is sputtering a crater, the second beam is progressively analyzing the crater bottom. The usual configuration for depth profiles is to sputter 300 × 300 $\mu m^2$ from the sample surface and the analysis is then done inside (in the middle) of this area 84 × 84 $\mu m^2$. After careful calibration to known spectral lines, the data files were analyzed by SurfaceLab software provided by IonTOF.

The sheet resistance $R_s$ values were measured using a four-point probe instrument (T2001A, Ossila). To ensure statistically relevant $R_s$ values, sheet resistance was measured at ten different spots on each sample and averaged. Measurement at one spot included subsequent recording of 25 $R_s$ values, obtained as an average value from 32768 measurements.

## Results and Discussion

### BND characterization

Figure 1 shows HRTEM images of the O-BND obtained after the purification/oxidation acid treatment (Figure 1a) and after subsequent hydrogenation (H-BND; Figure 1b). The O-BNDs exhibit elongated shapes with lengths ranging from 4 to 8 nm and thicknesses ranging from 2 to 4 nm. Both O-BND and H-BND display single crystal structures and hydrogenation does not have any significant effect on the atomic structure of BND. The HRTEM images show a predominant 111 and 220 interplanar distances and clear faceting of BNDs.

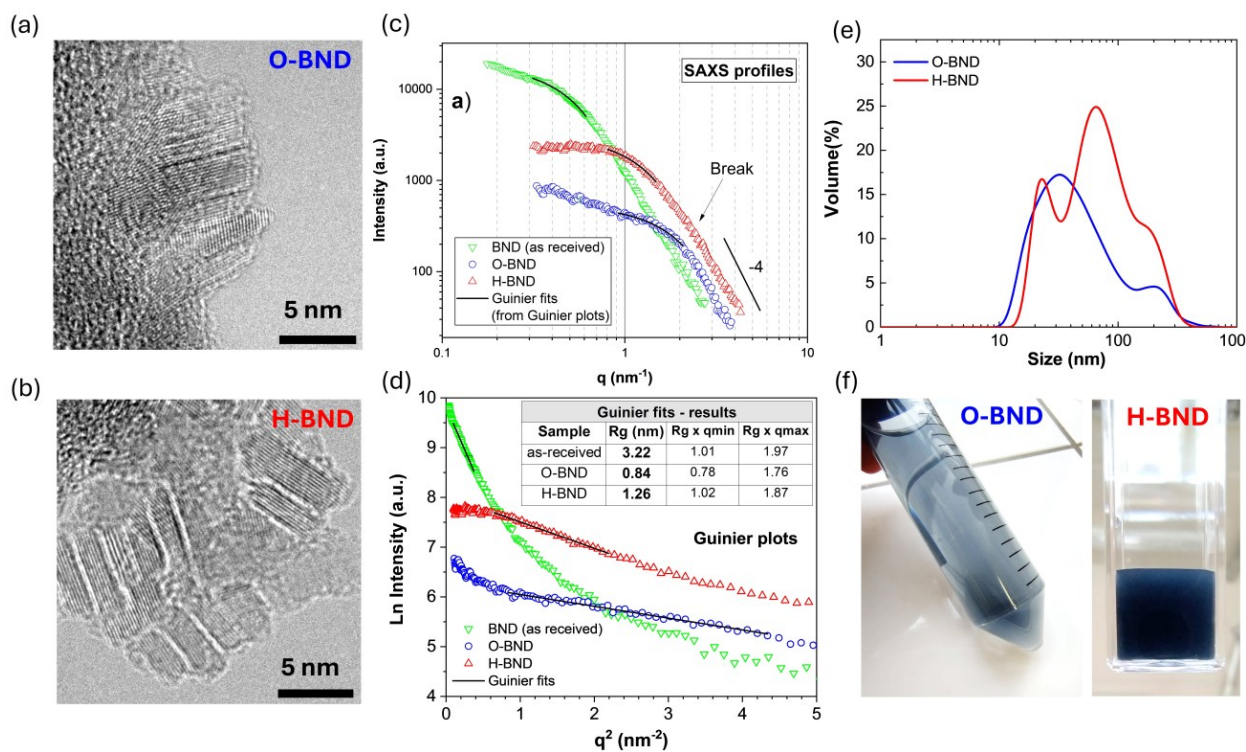

**Figure 1.** HRTEM images of O-BND (a) and H-BND (b). SAXS profiles of the as-received BND sample (green), O-BND (red) and H-BND (blue) with indicated Porod´s slope and break in the H-BND curve. DLS volume-weighted size distributions of O-BND (red) and H-BND (blue) (e). Photographs of the obtained colloidal solutions of O-BND (≈ 1 mg/ml) and H-BND (≈ 4 mg/ml) with a negative (-32 mV) and positive (+44 mV) zeta potential, respectively.

The SAXS profiles (Fig. 1c) of the as-received BND sample (green), the oxidized sample, O-BND (red), and the subsequently hydrogenated sample, H-BND (blue), clearly show the relative $q$ shifts of the respective SAXS knees, indicating changes in the nanoparticle or nanoaggregate sizes. Oxidation resulted in a significant shift of the "SAXS knee" to higher $q$ values, indicating that the particle size decreased or that the tightly bound clusters of ND crystallites were disaggregated. After subsequent hydrogenation, the particle size increased compared to the oxidized sample but remained smaller compared to the as-received sample.

The particle size (or cluster size) values, specifically their gyration radii ($R_g$), were evaluated by the Guinier analysis [52] using the *Simple Fits and Analysis* tool in *Irena* software. The Guinier plots, together with the respective Guinier fits, are shown in Fig. 1d. The inset table in Fig. 1d shows the resulting $R_g$ values for all samples together with the corresponding ranges of $R_g q$ products, demonstrating appropriate $q$ range selections for Guinier fits. The Guinier fits from Fig. 1d are also shown in the original SAXS profiles in Fig. 1c in order to illustrate the portion of the SAXS profile from which the $R_g$ was evaluated.

The $R_g$ values for ND particles in the as-received BND, O-BND and H-BND samples were 3.2 nm, 0.84 nm and 1.3 nm, respectively. For all samples, the fit could not be performed starting from the smallest $q$ values because the Guinier plots diverged from the linearity in the low-$q$ region. The as-received BND and O-BND samples exhibited a small-angle upturn in the low-$q$ region, potentially due to the presence of a limited number of larger particles or elongated particle shapes. H-BND sample showed a negative divergence from the linearity in the low-*q* region, indicating the presence of a structure factor – inter-particle interference of the scattered radiation. This structure-factor effect may lead to an underestimation of the nanoparticle size[53].

The high-*q* part of all SAXS profiles obeys the Porod's law ($I \sim q^{-4}$) relatively well, exhibiting a linear dependence with a slope of -4 in a double logarithmic scale. This indicates nanoparticles with a sharp interface between the particles and the surrounding matrix.

Independently of the exact nanoparticle morphologies, SAXS analysis clearly demonstrated that the oxidation process resulted in decreased particle size or disaggregation of tightly bound clusters to individual BND particles, which partially re-aggregated after the subsequent hydrogenation. Assuming homogenous particles or clusters with a spherical shape, the obtained gyration radii of as-received BND, O-BND and H-BND samples would correspond to particle/cluster dimensions of 8.3 nm, 2.2 nm and 3.3 nm, respectively. Another model fit (not presented here) using the *Unified Fit* tool in *Irena* software, taking into account the structure factor, provided an $R_g$ value of 1.6 nm corresponding to the particle size of 4.2 nm for H-BND sample. Noticeable break in the SAXS profile of the H-BND sample at the same location where the O-BND SAXS knee begins (at a *q* of about 2.5 nm$^{-1}$) suggests that the primary ND particles (crystallites) present in O-BND (with dimension around 2 nm) likely aggregated after the hydrogenation. The break at 2.5 nm$^{-1}$ in the H-BND profile corresponds to the primary ND particles forming the aggregates.

After ultrasonic dispersion in water, both O-BND and H-BND form stable colloids. The volume-weighted DLS data of the colloidal O-BND and H-BND solutions are shown in Figure 1e. The O-BND sample exhibits excellent dispersibility; the size spans from 10 to 300 nm with a mode value of 30 nm. After hydrogenation, the dispersibility becomes slightly worse with only a smaller portion of the H-BND being around 20 nm, and the maximum of the volumetrically dominant fraction around 65 nm. Figure 1f shows the intense blue color of both O-DND (1 mg/ml) and H-BND (4 mg/ml) colloidal dispersions, characteristic of highly boron-doped NDs[45,47]. The zeta potential of the O-BND was negative (-32 mV) at mildly acidic pH = 4-5 as a result of the oxidation process which results in the formation of abundant oxygen-rich, acidic groups. After hydrogenation in the hydrogen gas, the zeta potential became positive (+44 mV) at pH 5-6, as it is typical for hydrogenated nanodiamonds[54]. The surface chemistry and its evolution after hydrogenation was studied in detail by FTIR.

Figure 2 shows FTIR (a) and Raman (b) spectra of the O-BND and H-BND samples. After oxidation, the FTIR spectrum of the O-BND sample exhibits spectral features characteristic of oxidized NDs: a broad OH stretching band (2900 – 3700 cm$^{-1}$) accompanied by OH bending (1620 cm$^{-1}$) originating from the surface OH bonds and/or the adsorbed water[55], a strong C=O peak at 1760 cm$^{-1}$ and 1060 cm$^{-1}$ peak assigned to C-O bonds. In contrast to the typically observed features, the O-BND sample contains a sharp peak at 1220 cm$^{-1}$, tentatively assigned to B-C bonds arising from the boron incorporation in the diamond lattice.

Hydrogenation at 700°C greatly suppresses most of the oxygen-containing surface functional groups and introduces surface C-H$_x$ bonds. The C-H$_x$ stretching features are surprisingly sharp and contain only a few distinct peaks, unlike hydrogenated DND or HPHT ND[56]. Two prominent peaks can be readily assigned to C(111)-1x1-H plane stretching (2835 cm$^{-1}$) and to C(100):H stretching (2915 cm$^{-1}$)[55]. Such a sharp and "spectroscopically poor" region of C-H stretching indicates NDs with clear facets and rather uniform shape which is in agreement with the HRTEM observations (Figure 1a, b). The spectrum of the H-BND sample also contains the peak at 1220 cm$^{-1}$ which supports its assignment to the volume-related B-C bonds. The successful and efficient hydrogenation was accompanied by the zeta potential reversal from negative to positive as mentioned above.

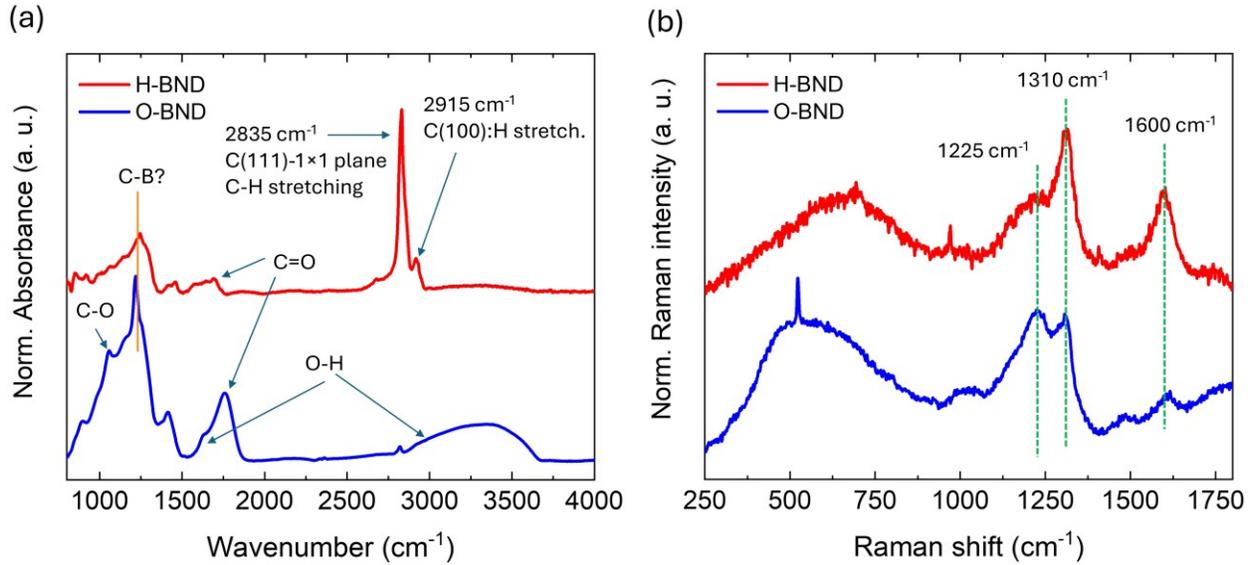

**Figure 2.** FTIR (a) and Raman (b) spectra of the O-BND (blue) and H-BND (red) samples.

Raman spectra of the O-BND and H-BND samples are shown in Figure 2b in blue and red, respectively. The Raman spectrum of O-BND exhibits features typical for highly boron-doped diamond[57]. First, the broadened and shifted peak at 1310 cm$^{-1}$ corresponds to the first-order T$_{2g}$ optical phonon mode (Zero-Center Phonon line, ZCP), originally peaking at 1332 cm$^{-1}$ in the perfect infinite diamond crystal. Second, an intense broad band at 1225 cm$^{-1}$ corresponds to a B$_2$ band of the diamond phonon density of states (PDoS)[58]. Finaly, a very broad band with a maximum of around 500 cm$^{-1}$ corresponds to an acoustic maximum of the PDoS peak (B$_1$ band)[58]. For a sample with high boron concentration, this effect is also connected with pronouncing of two boron-related wide bands at around 508 cm$^{-1}$ and 1205 cm$^{-1}$ (B$_1$ and B$_2$ bands, respectively) [59] that can be assigned to boron-dimer pairs and clusters vibrations and symmetry breaking of the diamond lattice, possibly due to boron-carbon complexes[59,60]. The spectrum of O-BND contains only traces of the non-diamond carbon that is manifested by the weak G-band at 1600 cm$^{-1}$ assigned to sp$^2$ carbon phases. Interestingly, after the hydrogenation, the Raman spectrum changed noticeably; the defect/boron-related features clearly decreased in intensity and the low-frequency broad band maximum shifted to higher wavenumbers (≈ 650 cm$^{-1}$). A similar effect was observed recently on the irradiated HPHT NDs[36] where the decrease of the defect-related features was explained by a healing effect of the high-temperature treatment on the lattice. In BNDs, where defects are probably substitutional boron atoms, the decrease of these features probably signalizes a partial diffusion of the boron out of the diamond lattice and thus a decrease of the boron concentration. Nevertheless, the H-BND sample keeps the characteristics of the BNDs, manifested, for example, by the intense blue color (Figure 1f). Notably, the non-diamond carbon content (G-peak at 1600 cm$^{-1}$) increased after the hydrogenation, which is consistent with previous reports for the HPHT ND and explained by the radical nature of the hydrogenation process[33,35]. Moreover, this increase is important for the positive zeta potential formation that is highly dependent on the non-diamond carbon content[32].

**Formation of seeding layers from H-BND, H-DND and TD_HPHT H-ND**

In order to investigate the effects of the particular ND characteristics on the electric properties of the grown BDD layer, seeding layers were prepared from three different NDs. All three NDs were hydrogenated which provided them with a positive zeta potential in water, which is an important prerequisite to form dense and uniform seeding layers on negatively charged Si or $SiO_2$ substrates[11,17,23,24]. Interestingly, the magnitude of the zeta potential differed among the three H-NDs. The zeta potential of H-BND and TD_HPHT H-ND was approximately 44 mV, H-DNDs were reported to achieve 52 mV at pH between 5-6, respectively[36]. It was recently proposed that the magnitude of the positive zeta potential of H-NDs correlates with the defect density, which is highest in the DND, explaining their highest zeta potential[36]. In contrast, the similarly prepared but intrinsic BU_HPHT NDs that are fully hydrogenated and with a minimal $sp^2$-C and defect content had a zeta potential of only 31 mV. Here, the H-BND showed 44 mV, which correlates with the non-negligible non-diamond content on the surface, as revealed by Raman spectroscopy (Figure 2b). It is also evident that boron incorporation does not inhibit the emergence of the positive zeta potential, although the boron effect on the magnitude and the long-term stability remains to be studied.

Further, there is a striking difference between the poor dispersibility of the intrinsic BU_HPHT ND prepared from halogenated adamantanes[38] and the BND prepared by the same BU_HPHT technique from 9BBN, which is perfectly dispersible after oxidation[45] as also demonstrated here. Obviously, the ND formation mechanism is different and highly dependent on the used molecular precursor. Here, we suppose that BNDs nucleate due to the polymerization of $sp^3$-hybridized molecules. From the analysis of the binding energies of atoms, it can be assumed that the most plausible polymerization mechanism is the crosslinking of neighboring molecules with B-C bonds involving tertiary carbon (the lowest energy of the C-H bond) and the release of molecular hydrogen. Then growth of the nanodiamond cluster becomes unstable. Tentatively we consider three stages of the 9BBN pyrolysis: i) polymerization with the formation of diamond-like clusters and boron-saturated nanographite, ii) growth of diamond clusters in nanographite, and iii) recrystallization of the nanophase mixture in a fluid growth medium. At the second stage (ii), excessive boron concentrates at the diamond surface, since diamonds do not accept all the boron from the nanographite phase with a composition of about BC8(9BBN = $C_{16}H_{30}B_2$). Probably, excessive boron hinders diffusion of carbon to diamond´s surface and this leads to samples with graphite phase between nanodiamonds. This interparticle graphite phase can then be selectively oxidized and the individual BND particles released. Such interparticle graphitic phase is missing in the intrinsic BU_HPHT NDs prepared from halogenated adamantanes and thus the prepared material is highly resistant to common oxidation treatments and has a poor dispersibility in water[36]. Thus, boron in 9BBN plays an important role not only as a dopant but also as an element enabling the formation of an interparticle graphitic phase, the selective removal of which by oxidation leads to the excellent dispersibility of BND.

Figure 3 shows the seeding layers formed on $SiO_2$ and $Si/SiO_x$ substrates for H-DND (a), TD_HPHT H-ND (b) and H-BND (c). Due to the insulating character of the $SiO_2$ substrate, we evidenced a strong charging of the H-DND/$SiO_2$ and partially also TD_HPHT H-ND/$SiO_2$ samples during SEM analyses. Therefore, we also used the conducting $Si/SiO_x$ substrates that enabled better inspection of the seeding layers of the particular H-NDs. Interestingly, the H-BND seeding layer was clearly observable even on insulating $SiO_2$ substrate indicating that H-BND were sufficiently electrically conductive to dissipate the charge acquired by the electron beam in SEM.

From the point of view of the seeding density, all the H-NDs formed relatively dense and homogeneous seeding layers. TD_HPHT H-ND and H-BND showed the highest surface coverage (25% and 24%, respectively), while the H-DND coverage was lower at 13%. For all the studied NDs, the seeding layer was formed by ND agglomerates rather than individual NDs. For H-BND, the size of the agglomerates corresponds well with the DLS data (Figure 1e), i.e. the seeding layer is predominantly formed by sub-100 nm fractions, yet some > 100 nm agglomerates can be identified.

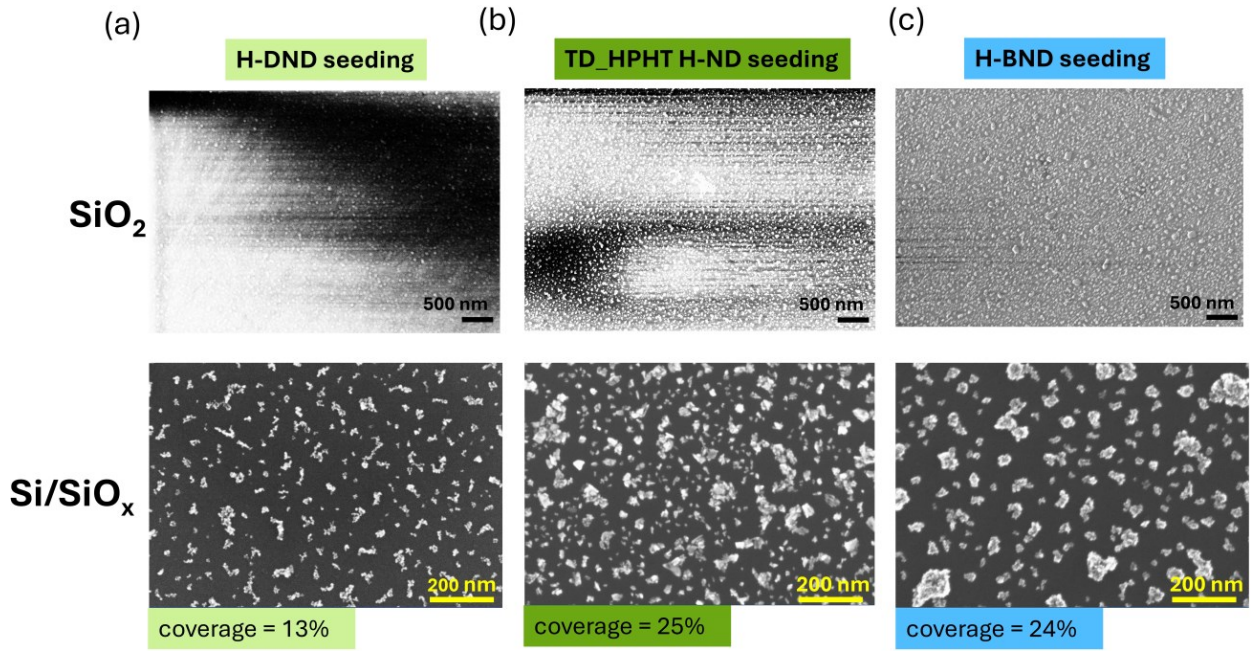

**Figure 3.** SEM images of seeding layers formed by H-DND (a), TD_HPHT H-ND (b) and H-BND (c) on $SiO_2$ (upper line) and on $Si/SiO_x$ (bottom line) substrate from the respective colloidal dispersion by a sonication treatment.

**BDD growth, morphology, and electrical properties**

The $SiO_2$ substrates seeded by three different H-NDs were finally used for the CVD growth of the BDD films. Figure 4 presents SEM images of the BDD films on H-DND (a), TD_HPHT H-ND (b) and H-BND (c) seeding layers. The thickness of the grown BDD films obtained by AFM was about 1 μm, as indicated above the SEM images in Figure 4. The similar thickness of all the BDD films indicates that the substrate coverage was not a critical parameter and even 13% coverage achieved with the H-DNDs was sufficient to grow a fully closed film with a thickness of nearly 1 μm.

The SEM images reveal clear morphological differences between the BDD films. The BDD film grown on the H-DND seeding layer consists of large (> 500 nm) as well as small (100-200 nm) crystals, with the large crystals often exhibiting defects such as stacking faults and multiple excessive twinning. In contrast, the BDD film grown on the TD_HPHT H-ND seeding layer consists of smaller (≤ 500 nm) but less defective crystals compared to the H-DND seeding layer. The relatively smaller crystal size in this film may be related to the largest number of individual NDs in the seeding layer. Finally, the film grown on the H-BND seeding consists of the largest (500-1000 nm) and least defective crystals.

To quantitatively assess the morphological differences, we performed boundary analysis of the SEM images, which provided a total length of the boundaries (interparticle and intraparticle) for a given sample. The processed images with identified grains and boundaries (highlighted red) are shown in the middle row of Figure 4. The bottom row depicts the calculated total boundary length for BDD films grown on H-DND, TD_HPHT H-ND, and H-BND seeding layers. Due to similar optical parameters (brightness and contrast) of interparticle (grain boundaries) and intraparticle (twin boundaries), it was not possible to distinguish the contribution of each particular type. A clear trend is observed, with the highest boundary length for H-DND seeding, a lower boundary length for TD_HPHT H-ND seeding, and the lowest for H-BND seeding.

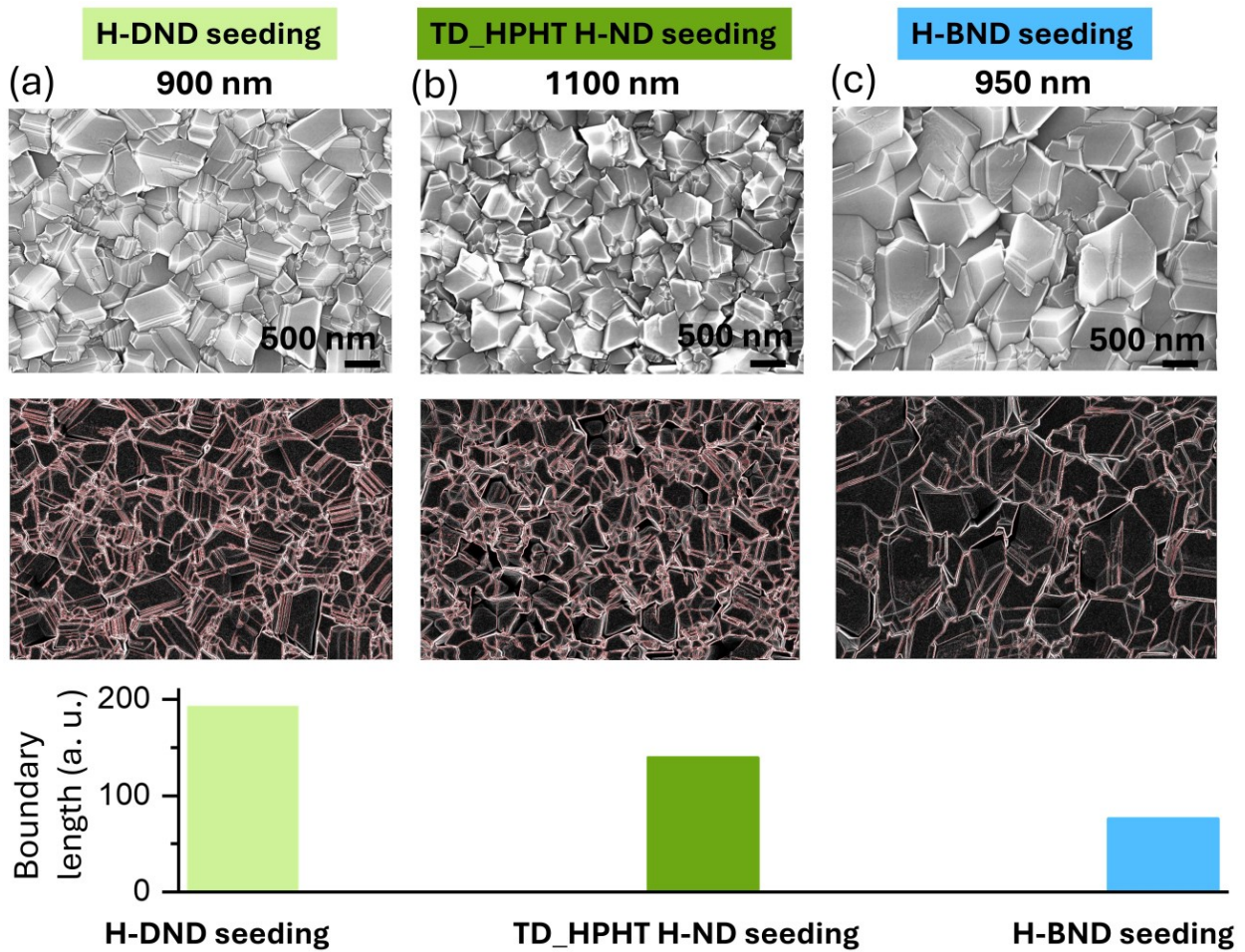

**Figure 4.** SEM images (upper row) of BDD films grown on the seeding layers formed by H-DND (a), TD_HPHT H-ND (b) and H-BND (c) on SiO$_2$ substrates. The middle row presents processed images with software-detected grain boundaries, and the bottom row shows the calculated boundary length for the BDD films grown on the H-DND, TD_HPHT H-ND and H-BND seeding layers.

The sheet resistance values of the BDD films, plotted in Figure 5a, were highest for films grown on the H-DND seeding and lowest for those grown on H-BND seeding. This trend correlates well with the observed morphological differences, particularly the boundary lengths depicted in Figure 4.

**Influence of Seeding Layer Properties on Morphology and Sheet Resistance**

The distinct morphologies observed in the BDD films can be attributed to two primary factors. The first obvious factor is the different intrinsic properties of the H-NDs used, such as size, crystal quality, and shape. Briefly, H-DNDs are intrinsically polycrystalline, defective nanodiamonds (~ 5 nm)[25] with excessive twinning. In contrast, the TD_HPHT H-NDs are single crystalline but with a wide size distribution (up to 30 nm) and irregular shapes[30]. While a wide size distribution may contribute to the large variations in grain size, the irregular shape could introduce additional defects. H-BNDs are single crystalline with very regular shapes and narrow size distribution, which is why they provide the least defective BDD film. From this perspective, the observed trend in morphology and defect density of the grown BDD films correlates well with the characteristics of respective H-NDs.

The second factor influencing film morphology is the seed density. Previous studies have shown that polycrystalline diamond film morphology can be controlled by variations in seeding density[61]. However, in the present work, the substrate coverage varied only by a factor of two between the H-NDs types (Figure 3), which is significantly lower one order of magnitude different seeding densities presented in the study of Vázquez-Cortés et al. [61]. In addition, the same morphological differences between BDD layers observed in Figure 4 were also observed on BDD layers grown on Si substrate (Figure S1) and $SiO_2$ substrate (Figure S2) with a lower B/C ratio, 3000 ppm. Therefore, it is likely that the dominant effect governing the different morphology is indeed seed quality rather than seeding density.

The SEM analysis (Figure 4) demonstrated a clear correlation between seed quality and the resulting crystal characteristics (size and defect concentration), indicating that the CVD growth of diamond on a seeding layer follows the homoepitaxial process. While interparticle grain boundaries are inevitable in the CVD-grown polycrystalline diamond film, we also observed significant intraparticle twin boundaries, notably in the BDD film grown of H-DND seeding (Figure 4a). Grain boundaries are known to reduce the electrical conductivity of BDD films by increasing the scattering of the free carriers; thus, the smaller the crystal size and longer the grain boundary length, the higher the resistance of the BDD film[62]. Grain boundaries also reduce the surface electrical conductivity in hydrogen-terminated diamond films. In our previous work focused on transport in hydrogen-terminated diamond films, we observed that the resistivity strongly increased when the mean grain size decreased from 400 to 40 nm while the Hall concentration was nearly the same[63].

The effect of twin boundaries on electronic properties has been thoroughly studied in the past for conventional semiconductors, such as Si[64] or GaAs[65], with effects varying according to the exact type of the semiconductor and the characteristics (orientation and atomic arrangement) of the twin boundaries. For example, it has been shown that coherent (111) - twin boundaries in silicon (the same lattice as diamond) have neither unusual diffusion properties nor influence on p-n junction behavior, and they appear indistinguishable from regular crystalline material[64]. Twining along (111) planes is a very common characteristic in DND[66], which consequently propagates into the CVD-grown BDD film (Figure 4a). However, a significant non-diamond C signal ($sp^2$ or distorted $sp^3$ coordinated carbon) was detected from both the DND coherent twin boundaries and the lamellar twinned regions. This indicates that even the coherent twin boundaries can deviate from their ideal symmetry and contain some non-$sp^3$ coordinated carbon[25]. Consequently, we propose that the twin boundaries also correspond to the free carriers' scattering and increase the BDD film's resistance. This would explain why the most defective BDD film, grown on the H-DND seeding layer, exhibited the highest sheet resistance, $R_s$ (Figure 5a).

BDD films grown on monocrystalline TD_HPHT H-NDs seeds/nuclei resulted in approx. 3 times $R_s$ compared to those grown on H-DND. This improvement can be attributed to the monocrystalline character of the nanoparticles, resulting in less defective CVD-grown crystals and reduced boundary length and, thus lower $R_s$ values. The lowest $R_s$ value was observed in BDD film grown on the H-BND seeds/nuclei, corresponding to the formation of the largest and least defective crystals with the shortest boundary length. Although the H-BND seeding layer primarily consisted of agglomerates rather than individual particles, a unique property of H-BND resulted in the growth of large and low-defect crystals.

The FTIR spectrum (Figure 2a) revealed that the surface C-$H_x$ bonds are predominantly located on the <111> and <100> facets which indicate a regular particle shape. Indeed, the TEM images (Figure 1a,b) further confirm a rather uniform shape of H-BND particles in contrast to the irregular and non-uniform morphology of TD_HPHT H-ND particles[30]. This morphological difference correlates with the much higher number of peaks in the C-$H_x$ stretching region reported for TD-HPHT H-ND[33,34,36]. Consequently, the most uniform particle shape and the monocrystalline character of the H-BND thus resulted in the lowest $R_s$ value.

Finally, the different qualities of H-NDs may influence the incorporation efficiencies of boron atoms into the grown BDD layer, which, in addition to different morphologies, could contribute to the observed $R_s$ variations, Figure 5a. Specifically, defects can act as preferential sites for boron incorporation, which can lead to a reduction in the doping efficiency in diamond crystals and a corresponding increase in $R_s$. To elucidate this issue, we performed SIMS analysis and compared the boron concentrations in BDD films grown on the different seeding layers. The depth profiling represented by time profiles of $B^-$ and $C^-$ ions for the BDD films grown from H-DND, TD_HPHT H-ND and H-BND seeding layers are shown in Figure 5b. The intensity profiles of $B^-$ and $C^-$ ions are almost identical for all three BDD films, confirming the same boron concentration in all three samples. These results show that the efficiency of boron incorporation was the same for all three films (and seeding layers) and finally prove the dominant effect of seed quality on the morphology of the resulting film and subsequently on the electronic properties, specifically the sheet resistance.

Figure 5a also shows the detailed trend in sheet resistance values as a function of surface termination. Systematically, higher sheet resistance values were measured on oxygen-terminated BDD layers of all types of seeding. This effect is possibly a combination of several factors: surface conductivity contribution induced by hydrogen termination[67] being also crystal size[63] and roughness-sensitive[68], a Schottky barrier[69] between the probes and oxidized BDD, and partial cleaning of the diamond surface layer (etching of conductive $sp^2$). Since the difference in $R_s$ between hydrogenated and oxidized BDD is only within the order of magnitude, as shown in Figure 5a, we assume that the main contributor to conductivity (sheet resistance) is the electroactive boron doping of the BDD film itself.

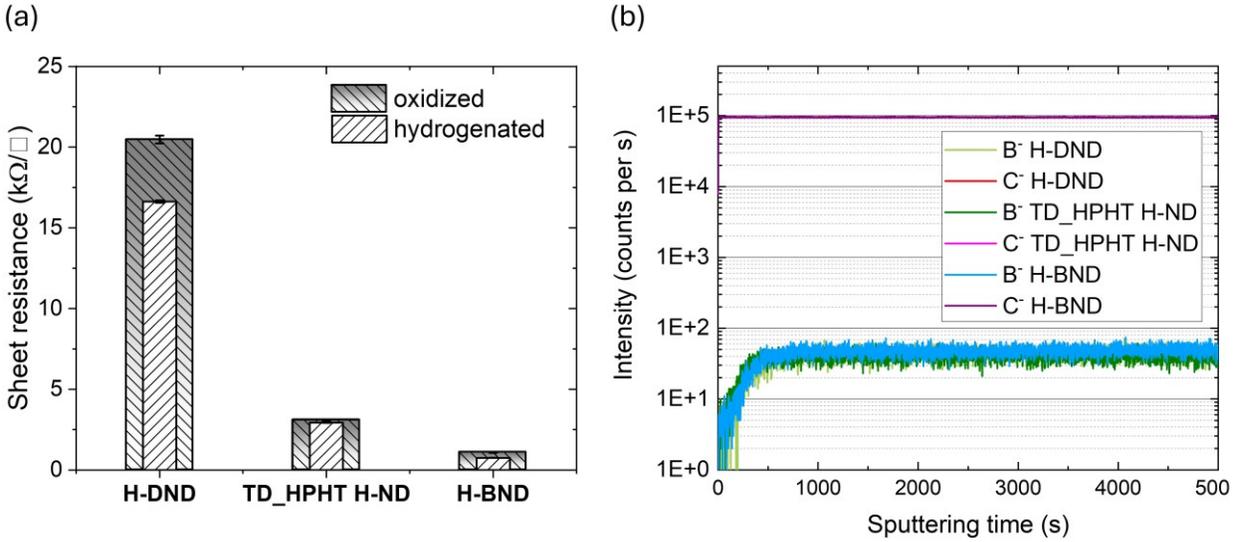

**Figure 5.** Sheet resistance values of BDD films grown on the seeding layers formed by H-DND, TD_HPHT H-ND and H-BND on SiO$_2$ substrates. Hydrogenated and oxidized values correspond to surface termination by H- and O- atoms (a). B- and C- ions time profile obtained by SIMS on the BDD films grown from H-DND, TD_HPHT H-ND and H-BND seeding layers (b).

The structure of the BDD films was finally studied by Raman spectroscopy in order to find a possible correlation with morphology. Raman spectra of BDD films grown on the different seeding layers are shown in Figure 6. All the Raman spectra exhibit characteristic features for boron-doped polycrystalline diamond material, namely broad bands peaking at 500 cm$^{-1}$ and 1200 cm$^{-1}$ and a zone-center-phonon (ZCP) peak at 1316-1319 cm$^{-1}$ corresponding to the first-order of the diamond optical phonon. The Raman spectra of all three samples are similar to each other, confirming the reproducibility of the growth process and the similarity of the obtained BDDs in terms of structure and boron doping already demonstrated by SIMS (Figure 5b). Nevertheless, we found minor differences in the G-band region (1500-1600 cm$^{-1}$) that correlate with morphology. All the samples exhibit two peaks in the G-band region. The G$_2$ band peaking around 1540 cm$^{-1}$ is characteristic of the nanocrystalline diamond films, indicating the presence of disordered sp$^2$-bonded material at the grain boundaries. The G$_1$ band probably comes from more isolated sp$^2$ defects[70]. The highest G$_1$ signal (≈ 1600 cm$^{-1}$) was observed for BDD film grown on H-DND. A strong signal at around 1600-1640 cm$^{-1}$ is characteristic of DND material[29] even after extensive oxidative purification[71] and probably comes from the isolated sp$^2$ defects that can appear on the surface and inside the particles[72], e.g., along the twin boundaries[25]. We tentatively assign the highest intensity of G$_1$ peak to the prominent content of the twin boundaries observed in BDD film grown on H-DND (Figure 4a). Correspondingly, the sp$^3$/sp$^2$ ratio was the lowest for the BDD films grown from the H-DND seeding layer and similar for the BDD films grown from TD_HPHT H-ND and H-BND seeding layers, as presented in Figure S3.

A long-term effort was devoted to finding a correlation between electric conductivity, boron content, and characteristic peaks in Raman spectra of boron-doped diamonds[59,73–76]. It was found that Fano-resonance multiple-curve model provides the best fits of BDD Raman spectra[59,73,74,76], although fitting using a combination of Gaussian and Lorentzian components also sufficiently expresses the trends

[75]. To roughly estimate boron concentration in diamond films from Raman spectra, we used a shift of the position and broadening of the ZCP peak, based on the model[73,74] designed for epitaxial diamond films. This approach is based on changes in the diamond peak parameters (position, width), reflecting defect density mostly assigned to boron[73]. Due to the presence of other defects like nitrogen, dislocations, and other non-sp$^3$ carbon phases, it has also been shown that boron concentration in the films itself does not directly correlate with their conductivity[73]. The carrier mobility is related to the maximum of electronic Raman scattering (peak at ca 500 cm$^{-1}$ and 350 cm$^{-1}$)[74].

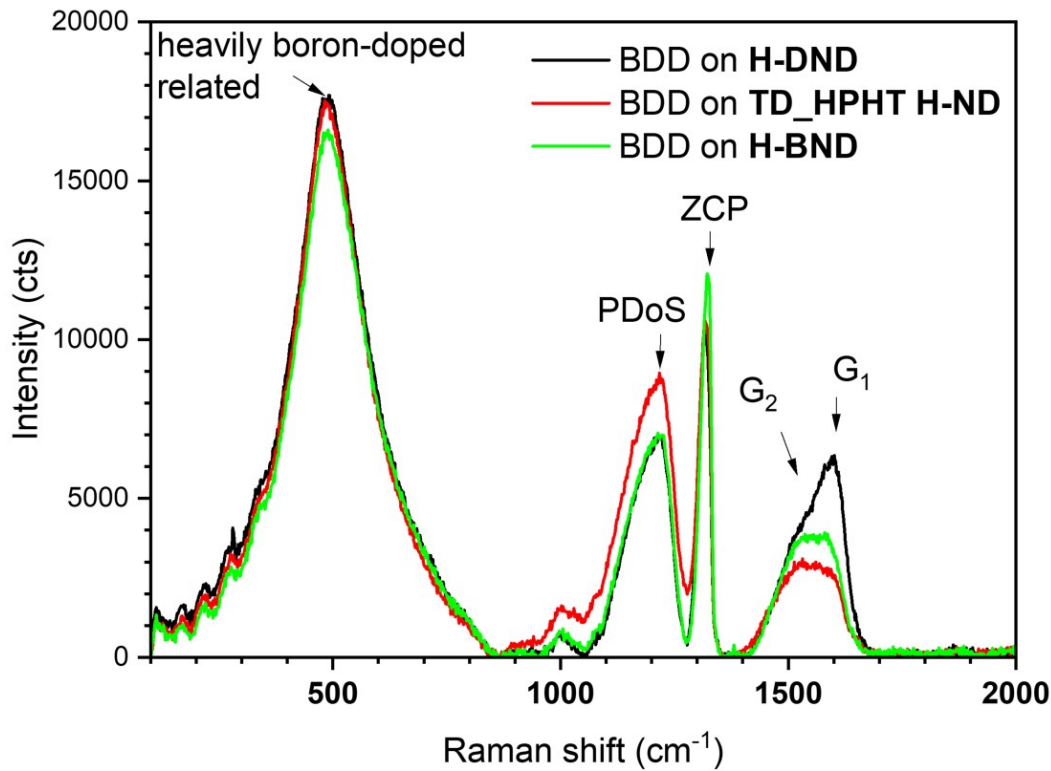

**Figure 6**. Raman spectra of BDD films grown on H-DND, TD_HPHT H-ND, and H-BND samples.

Based on these findings, we decided to express "defectiveness" of the samples structure qualitatively, i.e. to show trends. Contrary to the mentioned studies, we compared different films prepared under the same B/C ratio. Thus, we used a similar approach as Bernard et al.[75], i.e. we fitted Raman spectra with Gaussian functions with sufficient accuracy and analyzed ZCP peak parameters to estimate boron concentration as shown in Figure 7a. The BDD film grown on H-BND seeding layer shows the narrowest diamond peak (ZCP) as well as its closest position to the unperturbed diamond peak at 1332 cm$^{-1}$. On the contrary, the BDD film grown on the H-DND seeding layer exhibits the largest shift and FWHM of the ZCP peak. In between is the BDD film grown on the TD_HPHT H-ND layer. Since Raman spectroscopy provides information about the amount of defects in total, not only boron, for a qualitative estimation of the defect concentration we still compared the areas of Raman peaks having an origin other than the sp$^3$-phase. These results are shown in Figure 7b showing the same trends as in the case of the ZCP parameters. The results clearly correlate with the SEM morphology (Figure 4), i.e. confirming that the structure of BDD on H-BND is the least defective while the BDD on H-DND contains the highest defect concentration. Taking into account the results of Mortet and Taylor[73,74], the observed trends might be explained by increasing boron concentration from BDD on H-BND to BDD on the H-DND sample. However, this is not reflected in the $R_s$. Thus, we conclude that the concentration of boron is similar in all the samples, but the least defective structure (other defects than boron) of BDD on H-BND explains the lowest $R_s$ of this sample.

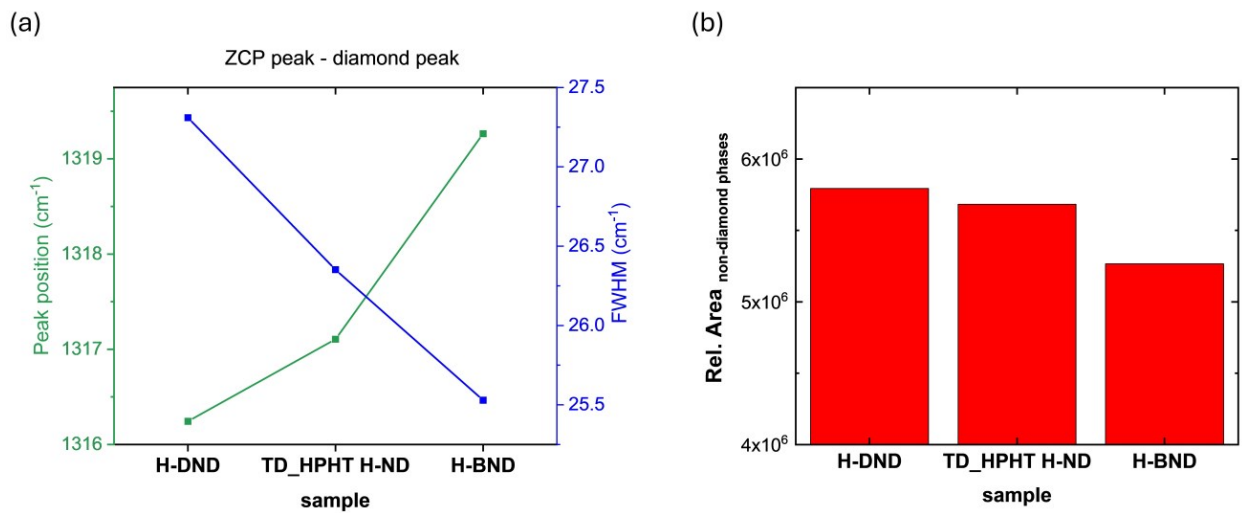

**Figure 7.** Position and halfwidth of ZCP peak for BDD grown on different seeding layers (a). Comparison of integral intensities of Raman peaks related to non-$sp^3$ phases to qualitatively express different defectiveness of the BDD structures grown on H-DND, TD_HPHT H-ND and H-BND seeding layers.

**Conclusion**

In this work, we employed novel boron-doped nanodiamonds (BND) synthesized by HPHT method from 9BBN organic precursor. After purification from $sp^2$-C content in oxidizing acids, clear, blue colloidal dispersions were obtained. SAXS, DLS and TEM characterization revealed negatively charged ($\zeta$ = -32 mV) well-dispersed O-BNDs in water, with a rather uniform, elongated shape that was apparent both in the TEM images and SAXS data. Raman spectroscopy confirmed a high boron doping level as well as excellent purity of the BNDs with minimal $sp^2$-C content. To reverse the zeta potential polarity, the O-BNDs were annealed in hydrogen at 700 °C. Similar to other hydrogenated nanodiamonds (e.g., DND), the hydrogenated BND (H-BND) had a positive zeta potential ($\zeta$ = +44 mV), demonstrating that boron doping does not prevent the acquisition of a positive zeta potential after hydrogenation. FTIR spectroscopy confirmed successful hydrogenation and highly uniform shapes of H-BNDs since only two dominant peaks in the C-$H_x$ stretching region of C(111)-1x1-H plane stretching (2835 $cm^{-1}$) and to C(100):H stretching (2915 $cm^{-1}$) were observed. The successful preparation of H-BNDs with positive zeta potential and extremely homogeneous shape allowed us to investigate the influence of the quality and type of nanodiamond seeding layer on the properties of BDD films. Along with H-BND we examined detonation nanodiamonds (H-DND) and bead-milled high-pressure high-temperature nanodiamonds (TD_HPHT H-ND). The seeding layers formed by these H-NDs exhibited varied surface coverages from 25% (TD_HPHT H-ND) and 24% (H-BND) down to 13% (H-DND). Despite these differences, all seeding layers supported the growth of fully closed ~ 1 μm thick BDD films. Although nearly identical boron concentrations detected by SIMS, significant differences in morphology and electrical properties were observed for BDD films grown on different seeding layers. The film grown on H-DND exhibited the highest crystal defects and sheet resistance, attributed to the polycrystalline and defective nature of primary DND particles. Improved quality with lower sheet resistance was achieved for TD_HPHT H-ND seeds. The BDD film grown on H-BND seeding demonstrated the lowest sheet resistance (440 ohm/sq), the largest crystals (500-1000nm), and the least defects. This superior performance is attributed to the uniform particle shape and monocrystalline character of H-BND, as evidenced by FTIR and TEM analyses.

These findings highlight the critical role of seeding layer properties, namely nanodiamond seeds' crystal quality and shape homogeneity, in determining the morphology and electrical characteristics of CVD-grown BDD films. The study underscores the potential of H-BND as an effective seeding material for high-quality BDD film growth, offering promising applications in electronic and electrochemical devices where low sheet resistance and high crystal quality are important, e.g., for thin diamond electrodes.

## Acknowledgement

The reported study was funded by RFBR (project number 20-52-26017) and by the Czech Science Foundation (GACR) (Grant No. 21-12567J). The work was also supported by Operational Programme Johannes Amos Comenius financed by European Structural and Investment Funds and the Czech Ministry of Education, Youth and Sports (Project No. SENDISO - CZ.02.01.01/00/22_008/0004596). We acknowledge CzechNanoLab Research Infrastructure supported by MEYS CR (LM2023051). S. S. acknowledges M. Pola, J. Libertinova and K. Hruska for technical support, sample treatments, and SEM analysis.

## References

[1] A. Aleksov, M. Kubovic, M. Kasu, P. Schmid, D. Grobe, S. Ertl, M. Schreck, B. Stritzker, E. Kohn, Diamond-based electronics for RF applications, Diam. Relat. Mater. 13 (2004) 233–240. https://doi.org/10.1016/j.diamond.2003.11.090.
[2] P. Ariano, A. Lo Giudice, A. Marcantoni, E. Vittone, E. Carbone, D. Lovisolo, A diamond-based biosensor for the recording of neuronal activity, Biosens. Bioelectron. 24 (2009) 2046–2050. https://doi.org/10.1016/j.bios.2008.10.017.
[3] R. Bogdanowicz, J. Ryl, Structural and electrochemical heterogeneities of boron-doped diamond surfaces, Curr. Opin. Electrochem. 31 (2022) 100876. https://doi.org/10.1016/j.coelec.2021.100876.
[4] D. Luo, L. Wu, J. Zhi, Fabrication of Boron-Doped Diamond Nanorod Forest Electrodes and Their Application in Nonenzymatic Amperometric Glucose Biosensing, ACS Nano 3 (2009) 2121–2128. https://doi.org/10.1021/nn9003154.
[5] Y. Gong, W. Jia, B. Zhou, K. Zheng, J. Gao, Y. Wu, Y. Wang, S. Yu, Y. Xue, Y. Wu, Heavily boron-doped polycrystalline diamond films: Microstructure, chemical composition investigation and plasma in-situ diagnostics, Appl. Surf. Sci. 659 (2024) 159838. https://doi.org/10.1016/j.apsusc.2024.159838.
[6] Y. Gong, W. Jia, B. Zhou, K. Zheng, D. Ma, Z. Li, J. Gao, Y. Ma, H. Hei, S. Yu, Y. Xue, Y. Wu, Effect of boron doping levels on the microstructure and characteristics of high-quality boron-doped diamond electrodes prepared by MPCVD, Diam. Relat. Mater. 139 (2023) 110377. https://doi.org/10.1016/j.diamond.2023.110377.
[7] T. Tsubota, T. Fukui, T. Saito, K. Kusakabe, S. Morooka, H. Maeda, Surface morphology and electrical properties of boron-doped diamond films synthesized by microwave-assisted chemical

vapor deposition using trimethylboron on diamond (100) substrate, Diam. Relat. Mater. 9 (2000) 1362–1368. https://doi.org/10.1016/S0925-9635(00)00254-5.

[8] A. Kromka, O. Babchenko, S. Potocky, B. Rezek, A. Sveshnikov, P. Demo, T. Izak, M. Varga, Diamond nucleation and seeding techniques for tissue regeneration, in: Diam.-Based Mater. Biomed. Appl., Elsevier, 2013: pp. 206–255. https://doi.org/10.1533/9780857093516.2.206.

[9] T. Yoshikawa, V. Zuerbig, F. Gao, R. Hoffmann, C.E. Nebel, O. Ambacher, V. Lebedev, Appropriate Salt Concentration of Nanodiamond Colloids for Electrostatic Self-Assembly Seeding of Monosized Individual Diamond Nanoparticles on Silicon Dioxide Surfaces, Langmuir 31 (2015) 5319–5325. https://doi.org/10.1021/acs.langmuir.5b01060.

[10] E.J.W. Smith, A.H. Piracha, D. Field, J.W. Pomeroy, G.R. Mackenzie, Z. Abdallah, F.C.-P. Massabuau, A.M. Hinz, D.J. Wallis, R.A. Oliver, M. Kuball, P.W. May, Mixed-size diamond seeding for low-thermal-barrier growth of CVD diamond onto GaN and AlN, Carbon 167 (2020) 620–626. https://doi.org/10.1016/j.carbon.2020.05.050.

[11] S. Stehlik, M. Varga, P. Stenclova, L. Ondic, M. Ledinsky, J. Pangrac, O. Vanek, J. Lipov, A. Kromka, B. Rezek, Ultrathin Nanocrystalline Diamond Films with Silicon Vacancy Color Centers via Seeding by 2 nm Detonation Nanodiamonds, ACS Appl. Mater. Interfaces 9 (2017) 38842–38853. https://doi.org/10.1021/acsami.7b14436.

[12] M. Iwabuchi, K. Mizushima, M. Mizuno, Y. Kitagawara, Dependence of Epitaxial Layer Defect Morphology on Substrate Particle Contamination of Si Epitaxial Wafer, J. Electrochem. Soc. 147 (2000) 1199. https://doi.org/10.1149/1.1393336.

[13] I. Mizushima, M. Koike, T. Sato, K. Miyano, Y. Tsunashima, Mechanism of Defect Formation during Low-Temperature Si Epitaxy on Clean Si Substrate, 38 (1999).

[14] S. Mandal, Nucleation of diamond films on heterogeneous substrates: a review, RSC Adv. 11 (2021) 10159–10182. https://doi.org/10.1039/D1RA00397F.

[15] N. Nunn, M. Torelli, G. McGuire, O. Shenderova, Nanodiamond: A high impact nanomaterial, Curr. Opin. Solid State Mater. Sci. 21 (2017) 1–9. https://doi.org/10.1016/j.cossms.2016.06.008.

[16] A. Krueger, D. Lang, Functionality is Key: Recent Progress in the Surface Modification of Nanodiamond, Adv. Funct. Mater. 22 (2012) 890–906. https://doi.org/10.1002/adfm.201102670.

[17] J. Hees, A. Kriele, O.A. Williams, Electrostatic self-assembly of diamond nanoparticles, Chem. Phys. Lett. 509 (2011) 12–15. https://doi.org/10.1016/j.cplett.2011.04.083.

[18] J. Cervenka, D.W.M. Lau, N. Dontschuk, O. Shimoni, L. Silvestri, F. Ladouceur, S.G. Duvall, S. Prawer, Nucleation and Chemical Vapor Deposition Growth of Polycrystalline Diamond on Aluminum Nitride: Role of Surface Termination and Polarity, Cryst. Growth Des. 13 (2013) 3490–3497. https://doi.org/10.1021/cg400383t.

[19] S. Mandal, C. Yuan, F. Massabuau, J.W. Pomeroy, J. Cuenca, H. Bland, E. Thomas, D. Wallis, T. Batten, D. Morgan, R. Oliver, M. Kuball, O.A. Williams, Thick, Adherent Diamond Films on AlN with Low Thermal Barrier Resistance, ACS Appl. Mater. Interfaces 11 (2019) 40826–40834. https://doi.org/10.1021/acsami.9b13869.

[20] S. Mandal, H.A. Bland, J.A. Cuenca, M. Snowball, O.A. Williams, Superconducting boron doped nanocrystalline diamond on boron nitride ceramics, Nanoscale 11 (2019) 10266–10272. https://doi.org/10.1039/C9NR02729G.

[21] G. Degutis, P. Pobedinskas, H.-G. Boyen, W. Dexters, W. Janssen, S. Drijkoningen, A. Hardy, K. Haenen, M.K. Van Bael, Improved nanodiamond seeding on chromium by surface plasma pretreatment, Chem. Phys. Lett. 640 (2015) 50–54. https://doi.org/10.1016/j.cplett.2015.10.002.

[22] P. Pobedinskas, G. Degutis, W. Dexters, J. D'Haen, M.K. Van Bael, K. Haenen, Nanodiamond seeding on plasma-treated tantalum thin films and the role of surface contamination, Appl. Surf. Sci. 538 (2021) 148016. https://doi.org/10.1016/j.apsusc.2020.148016.

[23] A. Kromka, B. Rezek, Z. Remes, M. Michalka, M. Ledinsky, J. Zemek, J. Potmesil, M. Vanecek, Formation of Continuous Nanocrystalline Diamond Layers on Glass and Silicon at Low Temperatures, Chem. Vap. Depos. 14 (2008) 181–186. https://doi.org/10.1002/cvde.200706662.

[24] T. Yoshikawa, F. Gao, V. Zuerbig, C. Giese, C.E. Nebel, O. Ambacher, V. Lebedev, Pinhole-free ultra-thin nanocrystalline diamond film growth via electrostatic self-assembly seeding with increased salt concentration of nanodiamond colloids, Diam. Relat. Mater. 63 (2016) 103–107. https://doi.org/10.1016/j.diamond.2015.08.010.

[25] S. Turner, O. Shenderova, F. Da Pieve, Y. Lu, E. Yücelen, J. Verbeeck, D. Lamoen, G. Van Tendeloo, Aberration-corrected microscopy and spectroscopy analysis of pristine, nitrogen containing detonation nanodiamond: Microscopy and spectroscopy analysis of pristine, nitrogen containing DND, Phys. Status Solidi A 210 (2013) 1976–1984. https://doi.org/10.1002/pssa.201300315.

[26] T. Ando, M. Ishii, M. Kamo, Y. Sato, Thermal hydrogenation of diamond surfaces studied by diffuse reflectance Fourier-transform infrared, temperature-programmed desorption and laser Raman spectroscopy, J. Chem. Soc. Faraday Trans. 89 (1993) 1783. https://doi.org/10.1039/ft9938901783.

[27] A.-I.- Ahmed, S. Mandal, L. Gines, O.A. Williams, C.-L. Cheng, Low temperature catalytic reactivity of nanodiamond in molecular hydrogen, Carbon 110 (2016) 438–442. https://doi.org/10.1016/j.carbon.2016.09.019.

[28] D. Miliaieva, A.S. Djoumessi, J. Čermák, K. Kolářová, M. Schaal, F. Otto, E. Shagieva, O. Romanyuk, J. Pangrác, J. Kuliček, V. Nádaždy, Š. Stehlík, A. Kromka, H. Hoppe, B. Rezek, Absolute energy levels in nanodiamonds of different origins and surface chemistries, Nanoscale Adv. 5 (2023) 4402–4414. https://doi.org/10.1039/D3NA00205E.

[29] S. Stehlik, M. Mermoux, B. Schummer, O. Vanek, K. Kolarova, P. Stenclova, A. Vlk, M. Ledinsky, R. Pfeifer, O. Romanyuk, I. Gordeev, F. Roussel-Dherbey, Z. Nemeckova, J. Henych, P. Bezdicka, A. Kromka, B. Rezek, Size Effects on Surface Chemistry and Raman Spectra of Sub-5 nm Oxidized High-Pressure High-Temperature and Detonation Nanodiamonds, J. Phys. Chem. C 125 (2021) 5647–5669. https://doi.org/10.1021/acs.jpcc.0c09190.

[30] I. Rehor, P. Cigler, Precise estimation of HPHT nanodiamond size distribution based on transmission electron microscopy image analysis, Diam. Relat. Mater. 46 (2014) 21–24. https://doi.org/10.1016/j.diamond.2014.04.002.

[31] O.A. Williams, J. Hees, C. Dieker, W. Jäger, L. Kirste, C.E. Nebel, Size-Dependent Reactivity of Diamond Nanoparticles, ACS Nano 4 (2010) 4824–4830. https://doi.org/10.1021/nn100748k.

[32] L. Ginés, S. Mandal, A.-I.-A. Ashek-I-Ahmed, C.-L. Cheng, M. Sow, O.A. Williams, Positive zeta potential of nanodiamonds, Nanoscale 9 (2017) 12549–12555. https://doi.org/10.1039/C7NR03200E.

[33] J. Mikesova, D. Miliaieva, P. Stenclova, M. Kindermann, T. Vuckova, M. Madlikova, M. Fabry, V. Veverka, J. Schimer, P. Krejci, S. Stehlik, P. Cigler, Nanodiamonds as traps for fibroblast growth factors: Parameters influencing the interaction, Carbon 195 (2022) 372–386. https://doi.org/10.1016/j.carbon.2022.04.017.

[34] L. Saoudi, H.A. Girard, E. Larquet, M. Mermoux, J. Leroy, J.-C. Arnault, Colloidal stability over months of highly crystalline high-pressure high-temperature hydrogenated nanodiamonds in water, Carbon 202 (2023) 438–449. https://doi.org/10.1016/j.carbon.2022.10.084.

[35] K. Kolarova, I. Bydzovska, O. Romanyuk, E. Shagieva, E. Ukraintsev, A. Kromka, B. Rezek, S. Stehlik, Hydrogenation of HPHT nanodiamonds and their nanoscale interaction with chitosan, Diam. Relat. Mater. 134 (2023) 109754. https://doi.org/10.1016/j.diamond.2023.109754.

[36] S. Stehlik, O. Szabo, E. Shagieva, D. Miliaieva, A. Kromka, Z. Nemeckova, J. Henych, J. Kozempel, E. Ekimov, B. Rezek, Electrical and colloidal properties of hydrogenated nanodiamonds: Effects of structure, composition and size, Carbon Trends 14 (2024) 100327. https://doi.org/10.1016/j.cartre.2024.100327.

[37] K. Henni, C. Njel, M. Frégnaux, D. Aureau, J.-S. Mérot, F. Fossard, I. Stenger, J.-C. Arnault, H.A. Girard, Core-shells particles grown in a tubular reactor: Influence of the seeds nature and MPCVD conditions on boron-doped diamond crystalline quality, Diam. Relat. Mater. 142 (2024) 110770. https://doi.org/10.1016/j.diamond.2023.110770.

[38] E.A. Ekimov, S.G. Lyapin, Yu.V. Grigoriev, I.P. Zibrov, K.M. Kondrina, Size-controllable synthesis of ultrasmall diamonds from halogenated adamantanes at high static pressure, Carbon 150 (2019) 436–438. https://doi.org/10.1016/j.carbon.2019.05.047.

[39] E.A. Ekimov, S.G. Lyapin, Yu.V. Grigor'ev, Carbonization of Brominated Adamantane and Nanodiamond Formation at High Pressures, Inorg. Mater. 56 (2020) 338–345. https://doi.org/10.1134/S0020168520030024.

[40] E.A. Ekimov, M.V. Kondrin, S.G. Lyapin, Yu.V. Grigoriev, A.A. Razgulov, V.S. Krivobok, S. Gierlotka, S. Stelmakh, High-pressure synthesis and optical properties of nanodiamonds obtained from halogenated adamantanes, Diam. Relat. Mater. 103 (2020) 107718. https://doi.org/10.1016/j.diamond.2020.107718.

[41] E.A. Ekimov, A.A. Shiryaev, V.A. Sidorov, Y.V. Grigoriev, A.A. Averin, M.V. Kondrin, Synthesis and properties of nanodiamonds produced by HPHT carbonization of 1-fluoroadamantane, Diam. Relat. Mater. 136 (2023) 109907. https://doi.org/10.1016/j.diamond.2023.109907.

[42] E. Ekimov, A.A. Shiryaev, Y. Grigoriev, A. Averin, E. Shagieva, S. Stehlik, M. Kondrin, Size-Dependent Thermal Stability and Optical Properties of Ultra-Small Nanodiamonds Synthesized under High Pressure, Nanomaterials 12 (2022) 351. https://doi.org/10.3390/nano12030351.

[43] E.A. Ekimov, M.V. Kondrin, V.S. Krivobok, A.A. Khomich, I.I. Vlasov, R.A. Khmelnitskiy, T. Iwasaki, M. Hatano, Effect of Si, Ge and Sn dopant elements on structure and photoluminescence of nano- and microdiamonds synthesized from organic compounds, Diam. Relat. Mater. 93 (2019) 75–83. https://doi.org/10.1016/j.diamond.2019.01.029.

[44] S.V. Bolshedvorskii, A.I. Zeleneev, V.V. Vorobyov, V.V. Soshenko, O.R. Rubinas, L.A. Zhulikov, P.A. Pivovarov, V.N. Sorokin, A.N. Smolyaninov, L.F. Kulikova, A.S. Garanina, S.G. Lyapin, V.N. Agafonov, R.E. Uzbekov, V.A. Davydov, A.V. Akimov, Single Silicon Vacancy Centers in 10 nm Diamonds for Quantum Information Applications, ACS Appl. Nano Mater. 2 (2019) 4765–4772. https://doi.org/10.1021/acsanm.9b00580.

[45] E.A. Ekimov, O.S. Kudryavtsev, A.A. Khomich, O.I. Lebedev, T.A. Dolenko, I.I. Vlasov, High-Pressure Synthesis of Boron-Doped Ultrasmall Diamonds from an Organic Compound, Adv. Mater. 27 (2015) 5518–5522. https://doi.org/10.1002/adma.201502672.

[46] A.M. Vervald, S.A. Burikov, A.M. Scherbakov, O.S. Kudryavtsev, N.A. Kalyagina, I.I. Vlasov, E.A. Ekimov, T.A. Dolenko, Boron-Doped Nanodiamonds as Anticancer Agents: En Route to Hyperthermia/Thermoablation Therapy, ACS Biomater. Sci. Eng. 6 (2020) 4446–4453. https://doi.org/10.1021/acsbiomaterials.0c00505.

[47] S. Heyer, W. Janssen, S. Turner, Y.-G. Lu, W.S. Yeap, J. Verbeeck, K. Haenen, A. Krueger, Toward Deep Blue Nano Hope Diamonds: Heavily Boron-Doped Diamond Nanoparticles, ACS Nano 8 (2014) 5757–5764. https://doi.org/10.1021/nn500573x.

[48] E.A. Ekimov, O.S. Kudryavtsev, S. Turner, S. Korneychuk, V.P. Sirotinkin, T.A. Dolenko, A.M. Vervald, I.I. Vlasov, The effect of molecular structure of organic compound on the direct high-pressure synthesis of boron-doped nanodiamond: Effect of organic compound on synthesis of boron-doped nanodiamond, Phys. Status Solidi A 213 (2016) 2582–2589. https://doi.org/10.1002/pssa.201600181.

[49] S. Stehlik, L. Ondic, M. Varga, J. Fait, A. Artemenko, T. Glatzel, A. Kromka, B. Rezek, Silicon-Vacancy Centers in Ultra-Thin Nanocrystalline Diamond Films, Micromachines 9 (2018) 281. https://doi.org/10.3390/mi9060281.

[50] J. Ilavsky, P.R. Jemian, *Irena* : tool suite for modeling and analysis of small-angle scattering, J. Appl. Crystallogr. 42 (2009) 347–353. https://doi.org/10.1107/S0021889809002222.

[51] G. Beaucage, Approximations Leading to a Unified Exponential/Power-Law Approach to Small-Angle Scattering, J. Appl. Crystallogr. 28 (1995) 717–728. https://doi.org/10.1107/S0021889895005292.

[52] T. Li, A.J. Senesi, B. Lee, Small Angle X-ray Scattering for Nanoparticle Research, Chem. Rev. 116 (2016) 11128–11180. https://doi.org/10.1021/acs.chemrev.5b00690.

[53] A.V. Smirnov, I.N. Deryabin, B.A. Fedorov, Small-angle scattering: the Guinier technique underestimates the size of hard globular particles due to the structure-factor effect, J. Appl. Crystallogr. 48 (2015) 1089–1093. https://doi.org/10.1107/S160057671501078X.

[54] V. Chakrapani, J.C. Angus, A.B. Anderson, S.D. Wolter, B.R. Stoner, G.U. Sumanasekera, Charge Transfer Equilibria Between Diamond and an Aqueous Oxygen Electrochemical Redox Couple, Science 318 (2007) 1424–1430. https://doi.org/10.1126/science.1148841.

[55] T. Petit, L. Puskar, FTIR spectroscopy of nanodiamonds: Methods and interpretation, Diam. Relat. Mater. 89 (2018) 52–66. https://doi.org/10.1016/j.diamond.2018.08.005.

[56] C.-L. Cheng, C.-F. Chen, W.-C. Shaio, D.-S. Tsai, K.-H. Chen, The CH stretching features on diamonds of different origins, Diam. Relat. Mater. 14 (2005) 1455–1462. https://doi.org/10.1016/j.diamond.2005.03.003.

[57] R. Jarošová, P.M. De Sousa Bezerra, C. Munson, G.M. Swain, Assessment of heterogeneous electron-transfer rate constants for soluble redox analytes at tetrahedral amorphous carbon, boron-doped diamond, and glassy carbon electrodes, Phys. Status Solidi A 213 (2016) 2087–2098. https://doi.org/10.1002/pssa.201600339.

[58] V. Mortet, A. Taylor, Z. Vlčková Živcová, D. Machon, O. Frank, P. Hubík, D. Tremouilles, L. Kavan, Analysis of heavily boron-doped diamond Raman spectrum, Diam. Relat. Mater. 88 (2018) 163–166. https://doi.org/10.1016/j.diamond.2018.07.013.

[59] V. Mortet, Z. Vlčková Živcová, A. Taylor, O. Frank, P. Hubík, D. Trémouilles, F. Jomard, J. Barjon, L. Kavan, Insight into boron-doped diamond Raman spectra characteristic features, Carbon 115 (2017) 279–284. https://doi.org/10.1016/j.carbon.2017.01.022.

[60] P. Knittel, R. Stach, T. Yoshikawa, L. Kirste, B. Mizaikoff, C. Kranz, C.E. Nebel, Characterisation of thin boron-doped diamond films using Raman spectroscopy and chemometrics, Anal. Methods 11 (2019) 582–586. https://doi.org/10.1039/C8AY02468E.

[61] D. Vázquez-Cortés, S.D. Janssens, E. Fried, Controlling the morphology of polycrystalline diamond films via seed density: Influence on grain size and film texture, Carbon 228 (2024) 119298. https://doi.org/10.1016/j.carbon.2024.119298.

[62] W. Gajewski, P. Achatz, O.A. Williams, K. Haenen, E. Bustarret, M. Stutzmann, J.A. Garrido, Electronic and optical properties of boron-doped nanocrystalline diamond films, Phys. Rev. B 79 (2009) 045206. https://doi.org/10.1103/PhysRevB.79.045206.

[63] P. Hubík, J.J. Mareš, H. Kozak, A. Kromka, B. Rezek, J. Krištofik, D. Kindl, Transport properties of hydrogen-terminated nanocrystalline diamond films, Diam. Relat. Mater. 24 (2012) 63–68. https://doi.org/10.1016/j.diamond.2011.10.021.

[64] H.J. Queisser, Properties of Twin Boundaries in Silicon, J. Electrochem. Soc. 110 (1963) 52. https://doi.org/10.1149/1.2425671.

[65] X. Qian, M. Kawai, H. Goto, J. Li, Effect of twin boundaries and structural polytypes on electron transport in GaAs, Comput. Mater. Sci. 108 (2015) 258–263. https://doi.org/10.1016/j.commatsci.2015.06.011.

[66] A.S. Barnard, G. Opletal, S.L.Y. Chang, Does Twinning Impact Structure/Property Relationships in Diamond Nanoparticles?, J. Phys. Chem. C 123 (2019) 11207–11215. https://doi.org/10.1021/acs.jpcc.9b00142.

[67] H. Kawarada, Diamond p-FETs using two-dimensional hole gas for high frequency and high voltage complementary circuits, J. Phys. Appl. Phys. 56 (2023) 053001. https://doi.org/10.1088/1361-6463/aca61c.

[68] O.A. Williams, R.B. Jackman, Surface conductivity on hydrogen terminated diamond, Semicond. Sci. Technol. 18 (2003) S34–S40. https://doi.org/10.1088/0268-1242/18/3/305.

[69] T. Teraji, Y. Koide, T. Ito, Schottky barrier height and thermal stability of p-diamond (100) Schottky interfaces, Thin Solid Films 557 (2014) 241–248. https://doi.org/10.1016/j.tsf.2013.11.132.

[70] A.C. Ferrari, J. Robertson, Raman spectroscopy of amorphous, nanostructured, diamond-like carbon, and nanodiamond, Philos. Trans. R. Soc. Math. Phys. Eng. Sci. 362 (2004) 2477–2512. https://doi.org/10.1098/rsta.2004.1452.

[71] S. Osswald, M. Havel, V. Mochalin, G. Yushin, Y. Gogotsi, Increase of nanodiamond crystal size by selective oxidation, Diam. Relat. Mater. 17 (2008) 1122–1126. https://doi.org/10.1016/j.diamond.2008.01.102.

[72] J.O. Orwa, K.W. Nugent, D.N. Jamieson, S. Prawer, Raman investigation of damage caused by deep ion implantation in diamond, Phys. Rev. B 62 (2000) 5461. https://doi.org/10.1103/PhysRevB.62.5461.

[73] A. Taylor, P. Ashcheulov, P. Hubík, Z. Weiss, L. Klimša, J. Kopeček, J. Hrabovsky, M. Veis, J. Lorinčík, I. Elantyev, V. Mortet, Comparative determination of atomic boron and carrier concentration in highly boron doped nano-crystalline diamond, Diam. Relat. Mater. 135 (2023) 109837. https://doi.org/10.1016/j.diamond.2023.109837.

[74] V. Mortet, I. Gregora, A. Taylor, N. Lambert, P. Ashcheulov, Z. Gedeonova, P. Hubik, New perspectives for heavily boron-doped diamond Raman spectrum analysis, Carbon 168 (2020) 319–327. https://doi.org/10.1016/j.carbon.2020.06.075.

[75] M. Bernard, A. Deneuville, P. Muret, Non-destructive determination of the boron concentration of heavily doped metallic diamond thin films from Raman spectroscopy, Diam. Relat. Mater. 13 (2004) 282–286. https://doi.org/10.1016/j.diamond.2003.10.051.

[76] F. Pruvost, A. Deneuville, Analysis of the Fano in diamond, Diam. Relat. Mater. 10 (2001) 531–535. https://doi.org/10.1016/S0925-9635(00)00378-2.

**Graphical abstract**

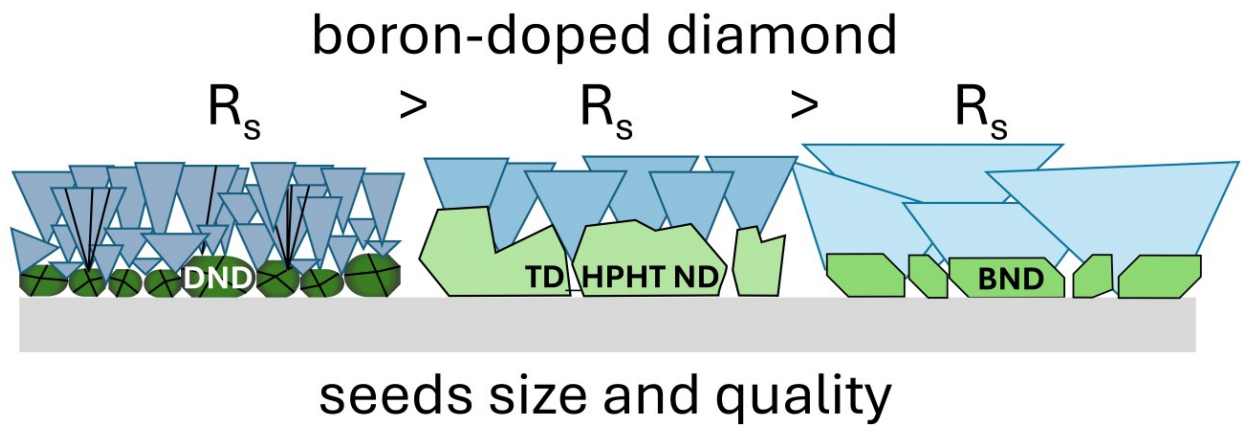

**Highlights:**

Hydrogenation and characterization of boron-doped nanodiamonds (BND) is presented.

Three types of hydrogenated nanodiamonds are used for seeding and compared.

Properties of boron-doped diamond films are studied in dependence on the seeding layer properties.

Defects and particle shape in the nanodiamond seeding layer strongly affect the BDD properties.

BND is a promising seeding material for high-quality BDD film growth.